# The Application of Cloud Computing to Astronomy: A Study of Cost and Performance

G. Bruce Berriman
Infrared Processing and Analysis Center
California Institute of Technology
Pasadena, California, USA
gbb@ipac.caltech.edu

Gideon Juve
Information Sciences Institute
University of Southern California
Marina del Rey, California, USA
gideon@isi.edu

Ewa Deelman
Information Sciences Institute
University of Southern California
Marina del Rey, California, USA
deelman@isi.edu

Moira Regelson
Raytheon Information Technology Services
Pasadena, California, USA
regelson@ipac.caltech.edu

Peter Plavchan
Infrared Processing and Analysis Center
California Institute of Technology
Pasadena, California, USA
plavchan@ipac.caltech.edu

***Abstract*—Cloud computing is a powerful new technology that is widely used in the business world. Recently, we have been investigating the benefits it offers to scientific computing. We have used three workflow applications to compare the performance of processing data on the Amazon EC2 cloud with the performance on the Abe high-performance cluster at the National Center for Supercomputing Applications (NCSA). We show that the Amazon EC2 cloud offers better performance and value for processor- and memory-limited applications than for I/O-bound applications. We provide an example of how the cloud is well suited to the generation of a science product: an atlas of periodograms for the 210,000 light curves released by the NASA Kepler Mission. This atlas will support the identification of periodic signals, including those due to transiting exoplanets, in the Kepler data sets.**

## I. Introduction

Vast quantities of data are made available to scientists at an ever-accelerating rate, and sophisticated and innovative approaches to data discovery, data mining and analysis are being developed to extract the full scientific content contained in this data tsunami. The e-Science paradigm is enabling the synthesis of new data products through the re-processing and re-sampling of existing data products. In this paper, we investigate the applicability of cloud computing to scientific applications. Cloud computing in this context refers to pay-as-you-go, on-demand compute resources made available by a third-party provider.

We study the cost and performance of one cloud service provider, Amazon EC2 (http://aws.amazon.com/ec2/), in running workflow applications. We investigate the performance of three workflow applications with different I/O, memory and CPU requirements on Amazon EC2, and compare the performance of the cloud with that of a typical high-performance cluster (HPC). Our goal is to identify which applications give best performance on the cloud at the lowest cost.

Finally, we describe the application of cloud computing to the generation of a new data product: an atlas of periodograms for the 210,000 light curves publicly released to date by the Kepler Mission. Kepler is designed to search for Earth-like exoplanets by observing their transits across their host star. The atlas of periodograms will support the identification of candidate exoplanets through the periodicities caused by the transits, as well as supporting studies of general variability in the Kepler data sets.

## II. Evaluating Applications on the Amazon EC2 Cloud

### A. Goals of This Study

Our goal is to determine which types of *scientific workflow applications* are cheaply and efficiently run on the Amazon EC2 cloud (hereafter, AmEC2). Workflows are loosely coupled parallel applications that consist of a set of computational tasks linked by data- and control-flow dependencies. Unlike tightly coupled applications, in which tasks communicate directly through the network, workflow tasks typically communicate using the file system: the output files written by one task become input files to be read by dependent tasks later in the workflow.

Given that AmEC2 uses only commodity hardware and given that applications make very different demands on resources, it is likely that cost and performance will vary dramatically by application. It was therefore important to study workflow applications that make different demands on resources. Thus the goals of this study are:

- Understand the performance of three workflow applications with different I/O, memory and CPU requirements on a commercial cloud
- Compare the performance of the cloud with that of a high-performance cluster (HPC) equipped with a high-performance network and parallel file system, and
- Analyze the various costs associated with running workflows on a commercial cloud.

## B. *Choice of Workflow Applications*

We have chosen three workflow applications because their usage of computational resources is very different: Montage (http://montage.ipac.caltech.edu) from astronomy, Broadband (http://scec.usc.edu/research/cme) from seismology, and Epigenome (http://epigenome.usc.edu) from biochemistry.

Montage [1] is a toolkit for aggregating astronomical images in Flexible Image Transport System (FITS) format into mosaics. Broadband generates and compares intensity measures of seismograms from several high- and low-frequency earthquake simulation codes. Epigenome maps short DNA segments collected using high-throughput gene sequencing machines to a previously constructed reference genome. Table I summarizes the relative resource usage of these three applications. The following three paragraphs give the technical specifications for the specific workflows used in this study.

TABLE I. SUMMARY OF RESOURCE USE BY THE WORKFLOW APPLICATIONS

| Application | I/O | Memory | CPU |
|---|---|---|---|
| Montage | High | Low | Low |
| Broadband | Medium | High | Medium |
| Epigenome | Low | Medium | High |

Montage was used to generate an 8-degree mosaic of M16 composed of images from the Two Micron All Sky Survey (2MASS; http://www.ipac.caltech.edu/2mass). The resulting workflow contained 10,429 tasks, read 4.2 GB of input data, and produced 7.9 GB of output data. Montage is considered I/O-bound because it spends more than 95% of its time waiting on I/O operations.

Broadband used four earthquake source descriptions and five sites to generate a workflow containing 320 tasks that read 6 GB of input data and wrote 160 MB of output data. Broadband is considered to be memory-limited because more than 75% of its runtime is consumed by tasks requiring more than 1 GB of physical memory.

The Epigenome workflow maps human DNA sequences to a reference copy of chromosome 21. The workflow contained 81 tasks, read 1.8 GB of input data, and produced 300 MB of output data. Epigenome is considered to be CPU-bound because it spends 99% of its runtime in the CPU and only 1% on I/O and other activities.

## C. *Experimental Set-Up*

In this section we summarize the experimental set-up. For a complete description, see [2] and [3]. We compared the performance of AmEC2 with that of the Abe High Performance Cluster (hereafter, Abe) at the National Center for Supercomputing Applications (NCSA; http://www.ncsa.illinois.edu/UserInfo/Resources/Hardware/Intel64Cluster), which is equipped with a high speed network and parallel file system to provide high-performance I/O.

To provide an unbiased comparison of the performance of workflows on AmEC2 and Abe, the experiments presented here were all run on single nodes, using the local disk on both AmEC2 and Abe. For comparison we also ran experiments using the parallel file system on Abe. Intuitively, the parallel file system would be expected to significantly improve the runtime of I/O-intensive applications like Montage, but would be less of an advantage for CPU-intensive applications like Epigenome.

Table II lists the five AmEC2 compute resources chosen for this study. We will refer to them by their AmEC2 designations *m1** and *c1**. These resources span the range of hardware offered by AmEC2, and include 32-bit and 64-bit machines, with memories from 1.7 GB to 15 GB, and from 1 to 8 cores. The *m1** resources use Opteron processors, while the *c1** resources use Xeon processors, which have the superior floating-point performance of the two (four floating-point operations per cycle vs. two floating-point operations per cycle).

The two Abe nodes use the same resource type—a 64-bit Xeon machine—but differ in the I/O devices used: *abe.local* uses a local partition for I/O, and *abe.lustre* uses a shared Lustre™ parallel file system. Both instances use a 10 Gbps InfiniBand™ network. The computational capacity of *abe.lustre* is roughly equivalent to that of *c1.xlarge*, which is useful when comparing the performance of Abe and AmEC2 and in estimating the virtualization overhead of AmEC2.

On AmEC2, executables were pre-installed in a Virtual Machine image, which is deployed on the node. The input data was stored in the Amazon Elastic Block Store (EBS) (a SAN-like storage service), while the output and intermediate files, as well as the application executables, were stored on local disks. For Abe, all application executables and input files were stored in the Lustre™ file system. For *abe.local* experiments, the input data were copied to a local disk before running the workflow, and all intermediate and output data were written to the same local disk. For *abe.lustre*, all intermediate and output data were written to the Lustre™ file system.

All jobs on both platforms were managed and executed through a job submission host at the Information Sciences Institute (ISI) using the Pegasus Workflow Management System (Pegasus WMS), which includes Pegasus [4] and Condor [5]. On AmEC2 we configured our VM image to start Condor worker processes when each node boots. The

TABLE II. SUMMARY OF THE PROCESSING RESOURCES ON AMAZON EC2 AND THE ABE HIGH-PERFORMANCE CLUSTER

| Type | Architecture | CPU | Cores | Memory | Network | Storage | Price |
|---|---|---|---|---|---|---|---|
| ***Amazon EC2*** | | | | | | | |
| *m1.small* | 32-bit | 2.0-2.6 GHz Opteron | 1-2 | 1.7 GB | 1 Gbps Ethernet | Local | $0.10/hr |
| *m1.large* | 64-bit | 2.0-2.6 GHz Opteron | 2 | 7.5 GB | 1 Gbps Ethernet | Local | $0.40/hr |
| *m1.xlarge* | 64-bit | 2.0-2.6 GHz Opteron | 4 | 15 GB | 1 Gbps Ethernet | Local | $0.80/hr |
| *c1.medium* | 32-bit | 2.33-2.66 GHz Xeon | 2 | 1.7 GB | 1 Gbps Ethernet | Local | $0.20/hr |
| *c1.xlarge* | 64-bit | 2.0-2.66 GHz Xeon | 8 | 7.5 GB | 1 Gbps Ethernet | Local | $0.80/hr |
| ***Abe Cluster*** | | | | | | | |
| *abe.local* | 64-bit | 2.33 GHz Xeon | 8 | 8 GB | 10Gbps InfiniBand | Local | N/A |
| *abe.lustre* | 64-bit | 2.33 GHz Xeon | 8 | 8 GB | 10 Gbps InfiniBand | Lustre™ | N/A |

Condor workers were configured to fetch application jobs directly from the submit host. On Abe we provisioned resources by submitting pilot jobs, using Globus GRAM, to the local PBS scheduler. The pilot jobs were used to start the Condor workers on Abe nodes. By using pilot jobs we were able to set up an execution environment on Abe that was equivalent to the one used on AmEC2.

## III. EVALUATING APPLICATIONS ON THE AMAZON EC2 CLOUD: RESULTS

### A. *Performance Comparison Between Amazon EC2 and the Abe High Performance Cluster*

Fig. 1 compares the runtimes of the Montage, Broadband and Epigenome workflows on all the Amazon EC2 and Abe platforms listed in Table II. Runtime in this context refers to the total amount of wall clock time, in seconds, from the moment the first workflow task is submitted until the last task completes. These runtimes exclude the following:

- The time required to install and boot the VM, which typically averages between 70 and 90 seconds (AmEC2 only)
- The latency in provisioning resources from Abe using the pilot jobs, which is highly dependent on the current system load (Abe only)
- The time to transfer input and output data, which varies with the load on the Wide Area Network (WAN). In our experiments, we typically observed bandwidth on the order of 500-1000KB/s between AmEC2 and the submit host in Marina del Rey, California.

This definition of runtime (also known as "makespan") enables a one-to-one comparison of the performances of AmEC2 and Abe. The similarities between the specifications of *c1.xlarge* and *abe.local* allow us to estimate the virtualization overhead for each application on AmEC2.

#### 1) *Montage (I/O-bound)*

The best performance was achieved on the *m1.xlarge* resource type. It has double the memory of the other types, and the extra memory is used by the Linux kernel for the file system buffer cache to reduce the amount of time the application spends waiting for I/O. This is particularly beneficial for an I/O-intensive application like Montage. Reasonably good performance was achieved on all resource types except *m1.small*, which is much less powerful than the other types. The AmEC2 *c1.xlarge* type is nearly equivalent to *abe.local* and delivered nearly equivalent performance (within 8%), indicating the virtualization overhead does not seriously degrade performance for this application.

The most important result of Fig. 1 is a demonstration of the performance advantage of high-performance parallel file systems for an I/O-bound application. While the AmEC2 instances are not prohibitively slow, the processing times on *abe.lustre* are nevertheless nearly three times faster than the fastest AmEC2 machines. A parallel file system could, in principle, be constructed from the raw components available in AmEC2, but without a high-speed network the I/O bottleneck would shift from the disk to the network. AmEC2 would need to deploy both a parallel file system and a high-speed interconnect to make dramatic performance upgrades. Recently Amazon released a new resource type that does include a 10Gb interconnect, however, we have not yet investigated its performance.

#### 2) *Broadband (Memory-bound)*

For Broadband the processing advantage of the parallel file system largely disappears: *abe.lustre* offers only slightly better performance than *abe.local*. And *abe.local*'s performance is only 1% better than *c1.xlarge*, so virtualization overhead is essentially negligible. For a memory-intensive application like Broadband, AmEC2 can achieve nearly the same performance as Abe as long as there is more than 1 GB of memory per core. If there is less, then some cores must sit idle to prevent the system from running out of memory or swapping. Broadband performs the worst on *m1.small* and *c1.medium*, the machines with the smallest memories (1.7 GB). This is because *m1.small* has only half a core, and *c1.medium* can only be configured to use one of its two cores because of memory limitations.

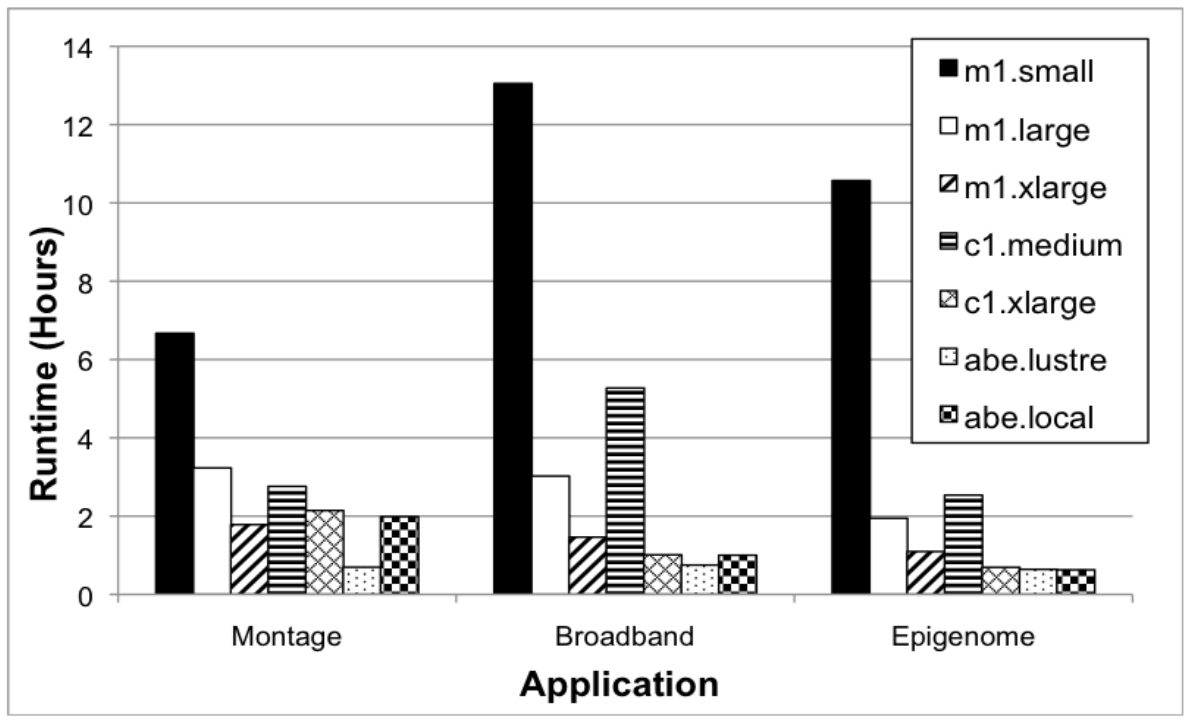

Figure 1. The processing times for the Montage, Broadband and Epigenome workflows on the Amazon EC2 cloud and the High Performance Cluster. The legend identifies the processors.

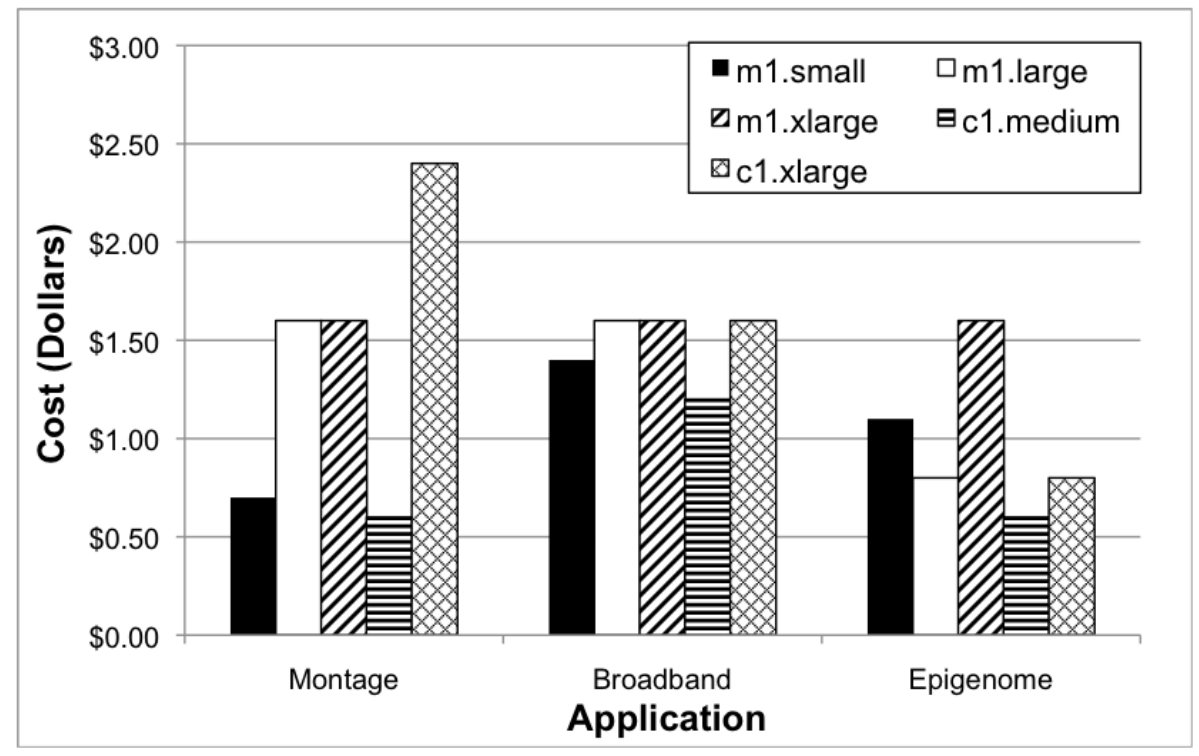

Figure 2. The processing costs for the Montage, Broadband and Epigenome workflows for the Amazon EC2 processors given in the legend.

*3) Epigenome (CPU-bound)*

As with Broadband, the parallel file system in Abe provides no processing advantage for Epigenome: processing times on *abe.lustre* were only 2% faster than on *abe.local*. Epigenome's performance suggests that virtualization overhead may be more significant for a CPU-bound application: the processing time for *c1.xlarge* was some 10% larger than for *abe.local*. The machines with the most cores gave the best performance for Epigenome, as would be expected for a CPU-bound application.

## IV. Cost-analysis of Running Workflow Applications on Amazon EC2

AmEC2 itemizes charges for the use of all of its resources, including charges for:

- Resources, including the use of VM instances and processing,
- Data storage, including the cost of virtual images in S3 and input data in S3,
- Data transfer, including charges for transferring input data into the cloud, and
- Transferring output data and log files between the submit host and AmEC2.

*1) Resource Cost*

Fig. 2 clearly shows the trade-off between performance and cost for Montage. The most powerful processor, *c1.xlarge*, offers a three-fold performance advantage over the least powerful, *m1.small*, but at five times the cost. The most cost-effective solution is *c1.medium*, which offers performance of only 20% less than *m1.xlarge* but at five-times lower cost.

For Broadband, the picture is quite different. Processing costs do not vary widely with machine, so there is no reason to choose less-powerful machines. Similar results apply to Epigenome**:** the machine offering the best performance, *c1.xlarge*, is also the second-cheapest machine.

*2) Storage Cost*

Storage cost is made up of the cost to store VM images in the Simple Storage Service (S3, an object-based storage system), and the cost of storing input data in the Elastic Block Store (EBS, a SAN-like block-based storage system). Both S3 and EBS use fixed monthly charges for the storage of data, and charges for accessing the data, which can vary according to the application. The rates for fixed charges are $0.15 per GB/month for S3, and $0.10 per GB/month for EBS. The main difference in cost is that EBS is charged based on the amount of disk storage requested, whereas S3 only charges for what is used. Additionally, EBS can be attached to only one computing instance, whereas S3 can be access concurrently by any number of instances. The variable charges for data storage are $0.01 per 1,000 PUT operations and $0.01 per 10,000 GET operations for S3, and $0.10 per million I/O operations for EBS.

The 32-bit image used for the experiments in this paper was 773 MB, compressed, and the 64-bit image was 729 MB, compressed, for a total fixed cost of $0.22 per month. The fixed monthly cost of storing input data for the three applications is shown in Table III. For the experiments described in this study, there were 4,616 S3 GET operations and 2,560 S3 PUT operations for a total variable cost of approximately $0.03. In addition, there were 3.18 million I/O operations on EBS for a total variable cost of $0.30.

TABLE III. Monthly storage Cost

| Application | Input Volume | Monthly Cost |
|---|---|---|
| Montage | 4.3 GB | $0.66 |
| Broadband | 4.1 GB | $0.66 |
| Epigenome | 1.8 GB | $0.26 |

*3) Transfer Cost*

In addition to resource and storage charges, AmEC2 charges $0.10 per GB for transfer into the cloud, and $0.17 per GB for transfer out of the cloud. Tables IV and V show the transfer sizes and costs for the three workflows. "Input" is the amount of input data to the workflow, "output" is the amount of output data, and "logs" refers to the amount of logging data recorded for workflow tasks and transferred back to the submit host. The cost of the protocol used by Condor to communicate between the submit host and the workers is not included, but it is estimated to be much less than $0.01 per workflow.

Table V summarizes the input and output sizes and costs. While data transfer costs for Epigenome and Broadband are small, for Montage they are larger than the processing and storage costs using the most cost-effective resource type. Given that scientists will almost certainly need to transfer data products out of the cloud, transfer costs may prove prohibitively expensive for high-volume applications. While the cost of transferring input data can be amortized by storing them in the cloud, the cost of transferring output data may be more difficult to reduce.

TABLE IV. DATA TRANSFER SIZES PER WORKFLOW ON AMAZON EC2

| Application | Input | Output | Logs |
|---|---|---|---|
| Montage | 4,291 MB | 7,970 MB | 40 MB |
| Broadband | 4,109 MB | 159 MB | 5.5 MB |
| Epigenome | 1,843 MB | 299 MB | 3.3 MB |

TABLE V. COSTS OF TRANSFERRING DATA INTO AND OUT OF THE AMAZON EC2 CLOUD

| Application | Input | Output | Logs | Total |
|---|---|---|---|---|
| Montage | $0.42 | $1.32 | $<0.01 | $1.75 |
| Broadband | $0.40 | $0.03 | $<0.01 | $0.43 |
| Epigenome | $0.18 | $0.05 | $<0.01 | $0.23 |

AmEC2 end-users of I/O-intensive applications producing large outputs need to give serious consideration to the trade-off between storage cost and transfer cost. Users may transfer input data for each run of a workflow, or transfer input data once, and store them in the cloud for multiple workflow runs. The choice of approach depends on how many times the data will be used, how long the data will be stored, and how frequently the data will be accessed. In general, storage is more cost-effective for input data that is reused often and accessed frequently, and transfer is more cost-effective if data will be used only once. For the workflows tested in this paper, the monthly cost to store input data is only slightly more than the cost to transfer it once. Therefore, for these applications, it may be more cost-effective to store the input data rather than transfer the data for each workflow, but a cost-benefit analysis should be performed in each case as part of selecting a processing and storage strategy.

### B. Sample Cost Effectiveness Study

We provide here a simple example of a cost-effectiveness study to answer the question: Is it cheaper to host an on-demand image mosaic service locally or on AmEC2? The costs described here are current as of October 2010. The calculations presented assume that the two services process requests for 36,000 mosaics of 2MASS images (total size 10TB) of size 4 sq deg over a period of three years. This workload is typical of the requests made to an existing image mosaic service hosted at the Infrared Processing and Analysis Center (http://hachi.ipac.caltech.edu:8080/montage/). Table VI summarizes the costs of the local service, using hardware choices typical of those used at IPAC. The roll-up of the power, cooling and administration are estimates provided by IPAC system management. Table VII gives similar calculations for AmEC2; the costs there include the costs of data transfer, I/O etc. Clearly, the local service is the least expensive choice. The high costs of data storage in AmEC2, and the high cost of data transfer and I/O in the case of an I/O-bound application like Montage, make AmEC2 much less attractive than a local service. An example of a much more cost-effective astronomy application will be given in Section V.

TABLE VI. COST PER MOSAIC OF A LOCALLY HOSTED IMAGE MOSAIC SERVICE

| Item | Cost ($) |
|---|---|
| 12 TB RAID 5 disk farm and enclosure (3 yr support) | 12,000 |
| Dell 2650 Xeon quad–core processor, 1 TB staging area | 5,000 |
| Power, cooling and administration | 6,000 |
| Total 3-year Cost | 23,000 |
| *Cost per mosaic* | 0.64 |

TABLE VII. COST PER MOSAIC OF A MOSAIC SERVICE HOSTED IN THE AMAZON EC2 CLOUD

| Item | Cost ($) |
|---|---|
| Network Transfer In | 1,000 |
| Data Storage on Elastic Block Storage | 36,000 |
| Processor Cost (*c1.medium*) | 4,500 |
| I/O operations | 7,000 |
| Network Transfer Out | 4,200 |
| Total 3-year Cost | 52,700 |
| *Cost per mosaic* | 1.46 |

## V. SUMMARY OF THE COMPARATIVE STUDY: WHEN TO USE THE CLOUD?

- Virtualization overhead on AmEC2 is generally small, but most evident for CPU-bound applications.
- The resources offered by AmEC2 are generally less powerful than those available in high-performance clusters and generally do not offer the same performance. This is particularly the case for I/O–bound applications, whose performance benefits greatly from the availability of parallel file systems. This advantage essentially disappears for CPU- and memory-bound applications.

- End-users should understand the resource usage of their applications and undertake a cost-benefit study of the resources offered to establish a processing and storage strategy. They should take into account factors such as:
  - Amazon EC2 itemizes charges for resource usage, data transfer and storage, and the impact of these costs should be evaluated.
  - For I/O-bound applications, the most expensive resources are not necessarily the most cost-effective.
  - Data transfer costs can exceed the processing costs for data-intensive applications**.**
  - Amazon EC2 offers no cost benefit over locally hosted storage, and is generally more expensive, but does eliminate local maintenance and energy costs, and does offer high-quality, reliable, storage.

## VI. Application to Calculation of Periodograms

### *A. The Need for High-Performance Periodogram Calculations*

The Kepler satellite (http://kepler.nasa.gov/), launched on 06 March 2009, is a NASA mission that uses high-precision photometry to search for transiting exoplanets around main sequence stars. The French mission Convection Rotation and Planetary Transits (CoRoT; http://www.esa.int/esaMI/COROT/index.html), launched in late 2006, has similar goals. Kepler's primary mission is to determine the frequency of Earth-sized planets around other stars. In May 2009, it began a photometric transit survey of 170,000 stars in a 105 square degree area in Cygnus. The photometric transit survey has a nominal mission lifetime of 3.5 years. As of this writing, the Kepler mission has released light curves of 210,664 stars; these light curves contain measurements made over 229 days, with between 500 to 50,000 epochs per light curve.

Analyzing these light curves to identify periodic signals, such as those that arise from transiting planets and from stellar variability, requires calculations of periodograms that reveal periodicities in time-series data and estimates of their significance. Periodograms are, however, computationally intensive, and the volume of data generated by Kepler demands high-performance processing. We have developed such a periodogram service, written in C, to take advantage of the "brute force" nature of periodograms and achieve the required performance. The processing of each frequency sampled in a periodogram is performed independently of all other frequencies, and so periodogram calculations are easily performed in parallel on a machine cluster by simply dividing the frequencies among the machines available. In practice, the processing is managed by a simple front-end job manager that splits the processing across all available machines, and then combines the results. The code itself returns the periodogram, a table of periodicities and their significance, light curves phased to the periodicities and plots of the periodograms and light curves.

The need for parallelization is shown in Table VI, which shows the processing times on a single Dell 1950 processor for three algorithms supported by the service.

TABLE VIII. Processing Times For Periodogram Algorithms on a Dell 1950 Server, with 2 x 2.5 GHz quad-core CPU's with 8 GB memory, running Red Hat Linux 5.3

| # Data Points | L-S | BLS | Plavchan | # Periods Sampled |
|---|---|---|---|---|
| 1,000 | 25 s | <15 s | 50 s | 100,000 |
| 10,000 | 5 min | 2 min | 14 min | 100,000 |
| 100,000 | 40 min | 15 min | 2 hr | 100,000 |
| 420,000 | 9 hr | 3 hr | 41 hr | 420,000 |

These algorithms are:

- **Lomb-Scargle (L-S).** Supports unevenly sampled data. Most useful for looking for sinusoidal-like variations, such as the radial velocity wobble of a star induced by an orbiting planet. [6, 7]
- **Box Least Squares (BLS)**. Optimized to identify "box"-like signals in time series data. Most useful for looking for transiting planets [8].
- **Plavchan.** Binless phase-dispersion minimization algorithm. It identifies periods with coherent phased light curves (i.e., least "dispersed"). There is no assumption about the underlying shape of the periodic signal. [9]

The processing times for light curves containing over 100,000 points, representative of the data sets that Kepler and CoRoT are expected to generate, can take well over an hour, and can require days in the case of the Plavchan algorithm. When run on a 128-node cluster of Dell 1950 processors, all the computations listed in Table VIII were sped-up by a factor of one hundred.

### *B. Calculating Periodograms On The Cloud*

To support the scientific analysis of Kepler data, we wished to generate an atlas of periodograms of the public Kepler data, computed with all three algorithms for maximal science value. The atlas will be served through the NASA Star and Exoplanet Database (NStED; http://nsted.ipac.caltech.edu), along with a catalog of the highest-probability periodicities culled from the atlas. End-users will be able to browse periodograms and phased light curves, identify stars for further study, and refine the periodogram calculations as needed.

We have computed the atlas on AmEC2, and there are several very good reasons for choosing it over a local cluster. The processing would interfere with operational services on the local machines accessible to us. The periodogram service has the characteristics that make it attractive for cloud processing. It is strongly CPU-bound, as it spends 90% of the runtime processing data, and the data sets are small, so the transfer and storage costs are not excessive. It is an example of bulk processing where the processors can be provisioned as needed and then released.

Table IX summarizes the results of a production run on the cloud. All 210,664 public light curves were processed with 128 processors working in parallel. Each algorithm was run with period sampling ranges of 0.04 days to 16.75 days and a fixed period increment of 0.001 days. The processing was performed in 26.8 hours, for a total cost of $303.06, with processing the major cost item at $291. The transfer cost is, however, significant because the code produced outputs of 76 GB—some four times the size of the input data.

The results showed that cloud computing is a powerful, cost-effective tool for bulk processing. On-demand provisioning is especially powerful and is a major advantage over grid facilities, where latency in scheduling jobs can increase the processing time dramatically.

TABLE IX. SUMMARY OF PERIODOGRAM CALCULATIONS ON THE AMAZON EC2 CLOUD

| | | **Result** |
|---|---|---|
| **Runtimes** | Tasks | 631,992 |
| | Mean Task Runtime | 6.34 sec |
| | Jobs | 25,401 |
| | Mean Job Runtime | 2.62 min |
| | Total CPU Time | 1,113 hr |
| | Total Wall Time | 26.8 hr |
| **Inputs** | Input Files | 210,664 |
| | Mean Input Size | 0.084 MB |
| | Total Input Size | 17.3 GB |
| **Outputs** | Output Files | 1,263,984 |
| | Mean Output Size | 0.124 MB |
| | Total Output Size | 76.52 GB |
| **Cost** | Compute Cost | $291.58 |
| | Transfer Cost | $11.48 |
| | **Total Cost** | **$303.06** |

## VII. CONCLUSIONS

Our study has shown that cloud computing offers a powerful and cost-effective new resource for scientists, especially for compute and memory intensive applications. For I/O-bound applications, however, high-performance clusters equipped with parallel file systems and high performance networks do offer superior performance. End-users should perform a cost-benefit study of cloud resources as part of their usage strategy.

We have used the calculation of an atlas of periodograms of light curves measured by the Kepler mission as an example of how the Amazon cloud can be used to generate a new science product. Although the monetary costs presented here were small, these costs can grow significantly as the number of curves grows, or as the search parameters are adjusted. As a result, commercial clouds may not be best suited for large-scale computations. On the other hand, there is now a movement towards providing academic clouds, such as those being built by FutureGrid (http://futuregrid.org/) or the National Energy Research Scientific Computing Center (NERSC) (http://www.nersc.gov/nusers/systems/magellan/) that will provide virtual environment capabilities to the scientific community. What remains to be seen is whether the level of service provided by academia can be on the par with that delivered by commercial entities.

## ACKNOWLEDGMENT

E. Deelman and G. Juve are funded by the National Science Foundation under Cooperative Agreement OCI-0438712 and grant # CCF-0725332. G. B. Berriman. and M. Regelson are supported by the NASA Exoplanet Science Institute at the Infrared Processing and Analysis Center, operated by the California Institute of Technology in coordination with the Jet Propulsion Laboratory (JPL).

## REFERENCES

[1] J. C. Jacob, D. S. Katz, G. B. Berriman, J. Good, A. C. Laity, E. Deelman, C. Kesselman, G. Singh, M.-H. Su, T. A. Prince, and R. Williams, "Montage: a grid portal and software toolkit for science-grade astronomical image mosaics." Int. J. Computational Science and Engineering. 2010. Vol, 4, Number 2, 1.

[2] G. B. Berriman, E. Deelman, P. Groth, and G. Juve. "The Application of Cloud Computing to the Creation of Image Mosaics and Management of Their Provenance": 2010. SPIE Conference 7740: Software and Cyberinfrastructure for Astronomy (GET REF) (2010)

[3] G. Juve, E. Deelman, K. Vahi, G. Mehta, G. B. Berriman, B. P. Berman, and P. Maechling, "Scientific Workflow Applications on Amazon EC2." 2010. Cloud Computing Workshop in Conjunction with e-Science Oxford, UK: IEEE . 2009

[4] Deelman, E. et al., "Pegasus: A framework for mapping complex scientific workflows onto distributed systems," Scientific Programming, 2005, 13, 219.

[5] Litzkow, M. J., et al., "Condor: A Hunter of Idle Workstations," 8th International Conference on Distributed Computing Systems, 1988, 104-111. 1988

[6] K. D. Horne and S. L. Baliunas, Ap J,1986. 302, 757

[7] J. D. Scargle. Ap J, 1982. 263,835

[8] G. Kovacs, S. Zucker, and T. Mazeh, A&A, 2002. 391, 377.

[9] P. Plavchan, M. Jura, J. D. Kirkpatrick, R. M. Cutri, and J. S. Gallagher. ApJS, 2008, 175,19.